\documentclass[12pt,preprint]{aastex}

\shorttitle{CHIPS observations of hot gas in Local Cavity}
\shortauthors{Hurwitz et al.}

\begin{document}


\title{Observations of Diffuse EUV Emission with the Cosmic
Hot Interstellar Plasma Spectrometer (CHIPS)}


\author{M. Hurwitz}
\affil{Space Sciences Laboratory, University of California,
    Berkeley, CA 94720}
\email{markh@ssl.berkeley.edu}

\author{T. P. Sasseen}
\affil{Department of Physics, UC Santa Barbara, Santa Barbara, CA, 93106}

\and

\author{M. M. Sirk}
\affil{Space Sciences Laboratory, University of California,
    Berkeley, CA 94720}



\begin{abstract}
The Cosmic Hot Interstellar Plasma Spectrometer (CHIPS) was designed 
to study diffuse emission from hot gas in the local interstellar cavity in the wavelength
range 90 - 265 \AA\/.  Between launch in January 2003 and early 2004, the instrument was operated
in narrow-slit mode, achieving a peak spectral resolution
of about 1.4 \AA\/ FWHM.  Observations were carried out preferentially
at high galactic latitudes;
weighted by observing time, the mean absolute value of the
galactic latitude for all narrow-slit observations combined is about 45 degrees.
The total integration time is about 13.2 Msec (74\% day, 26\% night).
In the context of a standard collisional ionization equilibrium plasma
model, the CHIPS data set tight constraints on
the emission measure at temperatures between 10$^{5.55}$ K and 10$^{6.4}$ K.
At 10$^{6.0}$ K, the 95\% upper limit on the emission measure is
about 0.0004 cm$^{-6}$ pc for solar abundance plasma
with foreground neutral hydrogen column of 2 x 10$^{18}$ cm$^{-2}$.
This constraint, derived primarily from limits on the extreme
ultraviolet emission lines of highly ionized iron, is well below 
the range for the local hot bubble estimated previously from soft X-ray studies.
If the pattern of elemental depletion in the hot gas follows
that observed in much denser interstellar clouds, the gas phase abundance 
of iron, relative to other heavy elements that contribute more to the soft 
X-ray emission, might be much lower than solar.
However, to support the emission measures inferred previously from
X-ray data would require depletions much higher than the moderate
values reported previously for hot gas.
Excluding the He {\sc II} Lyman
lines, which are known to be primarily terrestrial in origin, 
the brightest feature we find in the integrated spectrum 
is an Fe {\sc IX} line at 171.1 \AA\/.  
The sky-averaged flux of the feature is 
about 6 photons cm$^{-2}$ s$^{-1}$ ster$^{-1}$, a flux that exceeds
the 1-sigma shot noise significantly but is comparable to the 
systematic uncertainty.
We find "bright" 171.1 \AA\/ emission
(flux greater than 10 photons cm$^{-2}$ s$^{-1}$ ster$^{-1}$ and S/N $>$ 2)
in about 10\% of the observing time.  However, these "bright" observations 
overwhelmingly select for day time (96\% of 1.3 Msec).  Thus,
a local rather than interstellar origin for much of 
the 171.1 \AA\/ emission seems likely.

\end{abstract}




\section{Introduction}

Based on a number of methods of observation, the sun is known to lie
in a cavity of unusually low interstellar
density.  The cavity is believed to contain gas of temperature around 10$^6$ K with a 
filling factor greater than 80\%.  Denser material at the
cavity walls has been mapped in Na by \citet{Sfeir99}.
These measurements show the local cavity to extend
in the plane of the Galaxy about 100 pc, but much farther
perpendicular to the plane, possibly connecting to the lower
halo \citep{Crayfish02}. 

Analysis of the spatial and spectral distribution of the 
diffuse X-ray background
and HI shadowing reveal a relatively unabsorbed, "local" component 
with temperature about 10$^{6.1}$ K, emission measure 
ranging from about 0.0018 to 0.0058 cm$^{-6}$ pc, and a physical 
extent somewhat smaller than that of the boundary of the local cavity
\citep{Sno98,KS00,MS90}.
The collisional
ionization equilibrium (CIE) plasma emission codes employed in such
modeling predict a cluster of emission lines of highly
ionized iron (Fe {\sc VIII} - Fe {\sc XII}) near 72 eV, with a combined 
strength of about 150 to 300 photons cm$^{-2}$ s$^{-1}$ ster$^{-1}$ 
(hereafter "line units" or LU).
For a solar-abundance plasma in CIE near 10$^{6}$ K, the flux in the iron lines
comprises about 50\% of the total radiated power.  Thus
determining the strength of these lines observationally
has implications for plasma lifetimes.

Within the cavity exist warm, partially ionized
parcels of gas commonly called "clouds," 
in which the ionization fraction of helium is at or
above that of hydrogen despite helium's much higher ionization
potential \citep{Kimble93}.
Fossil nonequilibrium ionization can in principle provide an 
explanation for this puzzle, but the simplest scenario, in which
the clouds are recombining from a past impulsive event,
is difficult to reconcile with the abundance of neutral argon
in these clouds \citep{Jenkins00}.  For a solar-abundance
plasma in CIE near 10$^{6}$ K, the EUV iron lines provide about 90\% of the
helium-ionizing photons (weighting by the photoionization
cross-section).

Thus a picture has arisen in which nearby interstellar clouds are
photoionized, with stellar sources providing the bulk of
the hydrogen-ionizing photons, and the "local hot bubble" -- the 
hot medium itself, and/or its interfaces with cooler gas -- 
providing the bulk of the helium-ionizing photons \citep{VW95}.

Within this framework, several outstanding puzzles remain.
Interaction of the solar wind with the earth's exosphere \citep{Frey98}
and/or gas within the heliosphere \citep{lal04} may produce a substantial fraction
of the observed soft X-ray background, reducing
the need for a local hot interstellar component.  When the so-called "long term enhancements" 
in X-ray data are properly treated, the contribution
from earth's exosphere
can be limited to a small fraction of the observed signal \citep{CRS01},
but the remaining contributions are more difficult to constrain.
The few existing spectra of the X-ray background reveal thermal
line emission, but even an approximate fit to the data requires
depleted abundances and a complex temperature structure \citep{Setal01}.
Several observations, discussed below, have suggested that the energetically
important iron line complex near 72 eV is weaker than
expected, also potentially implicating abundance effects and/or
nonequilibrium ionization (the nonequilibrium model of \citet{BS94}, for
example, generally matches the broadband X-ray colors while producing
virtually no EUV line emission).
Thus, the key physical parameters such as temperature(s),
emission measure, elemental abundances, and ionization conditions
in the "local hot bubble" remain uncertain. 

The Cosmic Hot Interstellar Plasma Spectrometer (CHIPS), NASA's
first and to date only University-Class Explorer,
was designed to survey the sky for diffuse emission in the relatively unexplored 
band from about 90 to 265 \AA\/.  Of particular
interest is the cluster of emission features of Fe {\sc VIII} through Fe {\sc XII}
between about 168 and 195 \AA\/.  
Their close wavelength
spacing and common elemental origin makes them an excellent
thermometer.  
CHIPS' primary performance goal was to resolve
these features.  Subject to this resolution requirement, and the 
constraints imposed by the limited resources of the mission,
the instrument was designed to maximize
sensitivity near 170 \AA\/, and secondarily to include
as wide a bandpass as feasible.
In this paper we discuss data obtained in the first year or so
of the CHIPS mission and briefly describe the tests used to
verify the performance of the spectrograph on orbit.  

\section{Observations and Data Reduction} \label{obs}

CHIPS shares the orbit of ICESat,
the primary payload on the launch vehicle.
The orbit is approximately circular
at 600 km and an inclination of 94 degrees.
With all slits open, the spectrograph field of view is roughly
rectangular at 5 by 25 degrees.  The data presented in this work
were collected in narrow-slit mode, with a peak resolution
of about 1.4 \AA\/ FWHM.

We refer to a specific field on the sky (i.e., a certain
boresight and instrument roll angle) as a "target."
The selection of targets is influenced by
the demands of the spacecraft power and thermal
budgets and the capabilities of the attitude control system.
Within these constraints, we have sought to achieve deep
integrations on high galactic
latitude regions where, based on soft X-ray studies, the EUV emission was
expected to be brightest, while also sampling low and moderate latitudes.
Individual pointings are of $\sim$20 minute duration, 
typical of observatories in low-earth orbit.

The CHIPS spectrograph has six independent grating channels
dispersing diffuse emission across a common detector.
Thin-film filters attenuate out-of-band and scattered light.  
A diffuse emission line spans the full available detector
height, and most wavelengths land upon two distinct filters.
The emission features of iron are covered both by an aluminum
filter (the "large aluminum" panel) and a zirconium filter.
Additional details regarding the science instrument and satellite
are presented in \citet{hurwitz03}, \citet{janicik03}, \citet{marckwordt03}, \citet{sirk03},
and \citet{sholl03}.

We apply several filters to the raw data in producing
the reduced spectra.  Data are excluded when the overall
detector count-rate exceeds 80 events/sec, indicative of high charged-particle
backgrounds.  Data are also excluded when the detector high voltage is reduced
(a normal, temporary response to transient high-background periods).
A broad pulse-height filter is applied by the flight software.
Approximately 40\% of the telemetered events are then excluded
by a low-pass pulse-height threshold in ground software, selectively
reducing background because the low amplitude events are overwhelmingly
triggered by charged particles rather than by photons.

The data analysis pipeline automatically splits the photon
data from each target into day time and night time spectra.
We refer to each day or night spectrum as an "observation."
Because of the observing constraints mentioned above, there is
a strong correlation between a target's coordinates
and its availability in the orbital day/night cycle.  
The CY 2003-2004 narrow slit observing program resulted
in 244 targets and 355 observations. 

In Figure 1 we show the boresight locations, in galactic coordinates,
of the targets analyzed in this work.  
All narrow-slit data from CY 2003 and CY 2004 are included; each
target is indicated by a (+) symbol.
For targets in which either the day or night observation
resulted in an unusually high EUV flux (discussed below), 
a larger diamond surrounds the plotting symbol.
Weighted by observing time, the average absolute value of the
galactic latitude of the observed targets is about 45 degrees.

Periodically pulsed "stimulation pins" outside the active field of view are
used to register the event X,Y coordinates to a common frame,
thereby correcting for thermal drifts in the plate scale or
zero-points, and to determine the effective observing
time of the observation.  
Distortion is corrected using a preflight pinhole
grid map.  Small regions of known detector "hot spots" are
excluded, the event coordinates are rotated so that the new X axis
closely corresponds to the spectral dispersion direction, and
the spectra in each filter-half are summed over the active
detector height.

To determine the continuum against which to search
for emission lines, we use a very deep flight charged-particle spectrum,
scaled to match the local counts in the spectrum under analysis.  
The deep charged-particle spectrum generally contains many more photon events
per wavelength interval than does the spectrum under analysis,
and thus contributes relatively little to the shot noise
in the line flux.
Differences in the technique
by which the scale factor is determined -- for example, heavy smoothing
versus fitting low-order polynomials -- produce only $\sim$10\% changes
in the flux results.

Because it was expected that backgrounds would be
brighter than the photon signals of interest, we paid careful 
attention to systematic effects.
To determine the absolute "flatness" of the detector
response, we analyzed a a pre-flight photon 
flat field that was histogrammed like the flight spectra.
Compared to a smooth polynomial, the flat field spectrum
showed a pixel-to-pixel variance only slightly greater than
expected for shot noise, where "pixel"refers to the width of
the narrowest spectral features.  The excess   
one-sigma variance above shot noise was about 0.3\% of the total signal.  
Thus, in very deep spectra in which shot noise is negligible, 
excursions greater than 0.3\% compared to a locally smooth continuum are "real" 
at the one sigma level even without application 
of any flat field correction.

To explore the smoothness of the detector backgrounds in flight,
we applied an analysis similar to the one described above
to monthly "charged particle flat fields." 
We measured pixel to pixel excursions
relative not to a smooth polynomial
but to a scaled background formed by summing the charged particles
from the remaining months.  This analysis, like the analysis
applied to the flight photon spectra, includes a type of flat-field
correction.  We found a distribution of excursions closely
approximating a Poisson distribution, with no excess variance
over shot noise.
The monthly charged particle flat fields contain a total number of
counts corresponding to integration times of about 800,000 sec at
count-rates characteristic of our normal observing conditions.
Thus our analysis technique allows us to reach the shot
noise limit for ordinary spectra with integration times 
at least as deep as 800,000 sec.
We treat the shot noise in an 800,000 sec integration
as a limiting uncertainty for deeper observations, not because of 
evidence that systematic effects become significant at longer
times but because we have not yet shown their absence.

The wavelength scale is based on preflight measurements, offset by a 
(constant) $\sim$1 \AA\/
determined from the bright He {\sc II} 256.3 \AA\/
feature, which is present in essentially all the observations.  Temporal
variations in the measured centroid of the He {\sc II} feature shows a dispersion of about 0.2 \AA\/.  
Periodic observations of the moon, which provides a weakly reflected solar spectrum
that includes the features of Fe {\sc IX} through {\sc XII}, confirm that the adopted
wavelength scale provides a good fit near the center of the spectral
band, and that the relative throughput of the filter panels
is as measured pre-flight.  The measured fluxes of both the He {\sc II} feature
and the lunar spectrum are in good agreement with pre-flight expectations,
suggesting that the instrument throughput has not declined since laboratory
calibration.  Additional details of the on-orbit verification will be
presented in a subsequent work.

\section{Observational Results}

Over most of the CHIPS spectral band, and certainly near the wavelengths
of the iron lines, photoelectric absorption from the
neutral interstellar medium limits the observable path length
to at most a few hundred parsecs.  Thus the only astrophysical source we
expect to see is emission from the local hot bubble that
should be isotropic (within a factor of a few) on the sky.

In Figure 2 we show a spectrum from 150 - 200 \AA\/, binned at 0.5 \AA\/.
This is the complete spectrum for the CY 2003-2004 narrow slit program,
repesenting 13.2 Msec of effective observing time, inclusive
of all latencies or "dead times."  About 74\% of the data
were collected during orbital day; 26\% during orbital night.  
Both filter panels are included in the histogram.
Much of the observing time was directed
at regions where the EUV brightness was expected to be high based
on high soft X-ray brightness, low column density of neutral hydrogen, etc.
In Figure 2 we also show a scaled, charged-particle background spectrum
and an APEC/CHIANTI model (discussed below) with parameters from
one model for the local hot bubble (emission measure 0.0038 cm$^{-6}$ pc,
log(T)=6.1, foreground absorption 2 x 10$^{18}$ cm$^{-2}$) 
folded through the instrument response.
At most wavelengths, the small scale variations in the spectrum
are closely tracked by the charged-particle background, but there
is a modest excess of counts coincident
with the expected position of the Fe {\sc IX} line at 171.1 \AA\/.
For reference, the estimated strength of the observed line is about 6 LU.
The model prediction exceeds the measured flux in this line by
only a factor of a few.  However, essentially all of the other 
model lines arise from higher ionization stages of iron,
with predicted fluxes that
greatly exceed the limits set by the observations.

Individual observations can be identified in which the EUV emission
is brighter than in the integrated spectrum.  For this analysis,
we consider all observations with duration greater than 20 ksec.  
Of the 166 such observations, there are 11
in which the 171.1 \AA\/
feature is detected to a S/N (based on shot noise) greater than 2,
and for which the measured flux exceeds 10 LU.
Based on Gaussian statistics, only $\sim$4 observations should have 
exceeded a 2 sigma threshold if the true line flux were zero.  
The fact that 11 do indicates that
for most of these observations, the detection is not 
a statistical artifact.
Furthermore, examination of the
2-D detector images shows that the observed line is extended 
along the full slit length, ruling out a detector "hot spot." Thus,
despite the relatively low amplitude of
the feature even when it is bright, we are confident that it represents 
real photons entering the instrument through the narrow entrance slits.
As can be seen on Figure 1, the "bright" fields are generally
found at high latitudes, and there is a tantalizing grouping
near b = 60 degrees, l = 100 degrees.
But the most striking
feature of the bright observations is that only one of the 11 
occurred at night, and that observation was of relatively low
observing time (about 50 Ksec).  The overwhelming majority
of the bright observing time (96\%) corresponds to daytime data.
This greatly exceeds the daytime fraction for the observations
generally, and suggests that much of the emission is not interstellar
in origin but has a more local origin.  The flux of the 171.1 \AA\/ line 
in the combined spectrum of all the bright targets is 20 LU, with a 1 sigma
uncertainty (shot noise) of about 3 LU. 

Depending on the detailed gas temperature and elemental abundances,
a variety of features might potentially be detected
in the CHIPS spectra.
In Table 1, we show the flux at a variety of wavelengths
corresponding to bright single features or line clusters.
Results from both the complete data set and the EUV-bright observations
are presented.
To set the limits shown in Table 1, we compare the spectrum
to the scaled background over a wavelength interval 
1.4 times the FWHM resolution element at each stated wavelength.  
This interval is broad enough to capture essentially all
of the flux of a line at the stated wavelength, and allows us to
neglect the 0.2 \AA\/ uncertainty in the wavelength scale.
We also show the error in the mean flux (based on shot noise).
This shot noise error is small because the combined integration
time for all targets is 13,200,000 s.  But as discussed
above, we have verified independently that systematic
effects can be neglected for integration times of up to about 800,000 s,
not necessarily for 13,200,000 s.
Therefore in the the analysis below we take as the
1 sigma uncertainty in each line flux a value corresponding to the shot
noise limit for an $\sim$800,000 sec integration, and show this
uncertainty in Table 1.

The faintness of the EUV emission makes characterizing its
origin difficult.  Before launch we expected, or at least
hoped, that CHIPS would record signals of $\sim$50 LU or more.
Such signals could be detected to 3 sigma significance in
observing times of $\sim$200 Ksec.  The first year's
data set would provide nearly 70 independent spectra
at this sensitivity level, facilitating study of
how the EUV brightness might vary with astrophysical
parameters or with orbital
conditions (day/night cycle, zenith angle, sun avoidance, etc.).
In the actual data, the typical fluxes
are an order of magnitude fainter, demanding
integrations roughly 100 times deeper to achieve a given S/N.  
The narrow slit data set important limits on the brightness of 
EUV emission lines, and indicate that the periods when EUV
emission is detected are probably contaminated by a
foreground source associated with orbital day time.  
However, the narrow slit data
provide too few independent measurements
at sufficient sensitivity to allow the spatial / temporal / or
other variability of the emission to be explored in detail.
Accordingly, the instrument was switched to wide-slit
observing mode in early 2004.

\section{Collisional Ionization Equilibrium Plasma Modeling}

In this section we analyze the complete integrated spectrum, 
without excluding the EUV-bright observations or otherwise
correcting for the day time foreground emission.  This is
because the detailed foreground contribution is difficult
to quantify, and because, at the 95\% confidence level,
the constraints on plasma emission
measure are dominated by the limiting uncertainties rather than
the best-fit EUV line fluxes.  If the positive
best-fit flux(es) in the integrated spectrum were shown to arise 
entirely from the foreground, the constraints on plasma
emission measure would fall by at most a factor of 1.8
compared to those presented below.

To interpret the CHIPS data, we apply the predictions of the APEC
plasma code, utilizing a grid of line emissivities at temperature
steps of 0.1 dex provided by N. Brickhouse.  We also include 
features of Fe {\sc VIII} from the CHIANTI line list (version 4.2) \citep{Yetal03},
as these features are not present in the current release of APEC.
Emissivities are calculated in the low-density limit.

We generate a grid of model spectra with steps of 0.1
in the logarithm of the temperature and 0.00001 in the emission measure.
For each of the wavelengths at which we have a flux limit
in Table 1, we sum the fluxes of the nearby modeled
spectral lines,
weighting each by the appropriate value of
the resolution kernel.  We apply a foreground interstellar absorption with a 
column of 2 x 10$^{18}$ cm$^{-2}$,
compare the summed model line flux(es) with the observational limits set by
the integrated spectrum presented in Table 1,
calculate a value for chi-squared, and set
confidence intervals for two free parameters following \citet{LMB76}.

For gas with solar abundances, the CHIPS data constrain 
the plasma temperature
and emission measure as shown in Figure 3.
The best-fit parameters correspond
to a temperature around 10$^{5.8}$ K and an
emission measure of about 0.00014 cm$^{-6}$ pc.
The 68\% and 95\% confidence contours are illustrated
in Figure 3.  The central result is that, subject to the
model assumptions, the CHIPS data
constrain the emission measure tightly
for temperatures between about 10$^{5.55}$ K
and 10$^{6.4}$ K.
At 10$^{6.0}$ K, the 95\% upper limit is about 0.0004 cm$^{-6}$ pc.

For depletions in which the gas phase
iron abundance is reduced by a factor of only a few compared to solar,
depletion trades off directly with the allowed emission
measure.  This is because the modeled iron features are
significantly brighter all other features in the CHIPS band
unless iron is depleted significantly.  
The iron abundance inferred from an X-ray spectrum of 
local hot gas was about 30\% of solar \citep{Setal01};
for such gas, the constraints on emission measure
contained in Figure 3 should simply be scaled upward
by a factor of about 3.

To explore the case of highly depleted
gas, we have repeated the
analysis described above assuming "warm cloud" depletions from \citet{SS96}.
In this model, iron is depleted by 1.2 dex, or to a value
about 1/16 of solar, and other elements are depleted as well.
We show the corresponding constraints on plasma temperature
and emission measure in Figure 4.  

To explore the degree to which our results are dependent
on the iron lines, we constructed an artificial set of
depletions in which the gas phase iron abundance is zero,
and all other elements have solar abundances.  For that model,
the constraints on emission measure are
similar to those shown in Figure 4.  Thus, the
"warm cloud" depletions treated above already suppresses the 
iron sufficiently that the emission measure is limited by
nondetection of lines of other elements of relatively low depletion
(for example, a cluster of O {\sc V} and O {\sc VI} lines 
near 171 - 173 \AA\/).

\section{Discussion}

The CHIPS data show that the EUV iron lines are extremely faint,
and even the fluxes analyzed here are likely to include
a local / foreground component, leaving even less room for
interstellar flux.  
Although previous works have suggested that the Fe features are
weaker than predicted by collisional ionization equilibrium models, 
some of those results have rested on
uncertain and untestable assumptions.
Using a point-source spectrometer aboard
the Extreme Ultraviolet Explorer, \citet{JVE95} set an upper limit
to the emission measure near 10$^{5.8}$ K that is also lower
than expected for the local hot bubble (but not as
tight as our limits).  However, owing in part to the low resolution
of the instrument, it was difficult to confirm that the very large
background in the EUVE signal could be subtracted precisely.
\citet{Bloch02} interpret ALEXIS multilayer imaging of the
diffuse background with a model in which an astrophysical
EUV flux is assumed to scale with the Wisconsin B-band map,
and conclude that the Fe emission is no more than about 1/10
of the predicted flux level.  
The pulse-height distribution for the Be-band proportional counter Wisconsin 
rocket data has been reported to require significant depletions (i.e., the
pulse height distribution is inconsistent with full-strength iron lines) \citep{B88}. 
Using a rocket-borne calorimeter, \citet{MCC02} report a detection of emission
from Fe {\sc IX}, {\sc X}, and {\sc XI} (unresolved) with a combined flux 
of about 100 LU, significantly greater than
allowed by the CHIPS data.  The calorimeter signal (4 events) exceeds the 
expected background (0.3 events) and sets a 90\% lower limit of about 40 LU 
on the combined line strengths, a limit well above the best-fit CHIPS results.
Whether the difference in results arises from instrument
calibration effects, a statistical oddity, or genuine differences
in fields observed (due to astrophysical or local effects) is
not yet clear.  In any case, the CHIPS observations set robust constraints on
important individual spectral lines.

The CHIPS results are at variance with X-ray measurements interpreted
with similar models and subject to the same assumptions (solar abundances,
collisional ionization equilibrium).  As noted above, temperatures 
greater than 10$^{6.0}$ K, and emission measures an order of 
magnitude higher than the 95\% upper limits of Figure 3, have been reported.
Depletion of iron relative to other heavy elements generally
helps reconcile the EUV and X-ray results, but the depletions
reported previously have been only moderate: roughly a factor of 3
reduction in the gas phase abundance of elements including iron \citep{Setal01,MCC02}.  
The CHIPS results are qualitatively consistent with a 
poster presented by \citet{Bellm03},
concluding that a combination of depletion and a lower
temperature (10$^{5.85}$ K) in the local hot gas provides
a better fit to the broadband X-ray colors compared
to undepleted, hotter models.

Foreground absorption alone cannot reconcile the CHIPS and X-ray results.
A foreground column density of N{\sc HI} of 
1.5 x 10$^{19}$ cm$^{-2}$ is required in order for
the hydrogen and associated helium to attenuate
the iron features by about a factor of 10.  Such a foreground 
column would violate the observed constancy of the B-band to 
Be-band ratio \citep{J88}.  

We have not obtained a detailed charge-exchange spectrum
that can be compared directly with the CHIPS data.  We note
however that the solar wind (esp. the slow equatorial wind)
is enriched in Fe due to the well known "FIP" effect,
and that the dominant ionization stages are Fe {\sc X} and Fe {\sc XI}
\citep{SFZ99,vS94}.  Charge exchange provides at least
a plausible origin for the Fe {\sc IX} emission detected
in the bright observations.  The greater its
contribution to the observed EUV flux, of course, the tighter the constraints
on emission attributable to an interstellar plasma.

Reconciling in detail
the CHIPS observations with the existing body of X-ray data
on the local bubble will evidently require careful treatment 
of depletion effects.  Whether the models must also multi-temperature
structures or nonequilibrium interstellar conditions in the local
bubble is not yet known.






\acknowledgments

CHIPS is supported by NASA Grant NAG 5-5213.
We are grateful to the CHIPS hardware and flight operations teams, 
Mr. David Pierce, Dr. Randy Kimble, and the many review panel members
whose participation strengthened the project.
Nancy Brickhouse graciously provided the APEC EUV line emissivities
in a convenient format and, with Randall Smith, patiently
answered questions as to format and content.  
We utilized the data compiled by CHIANTI,
a collaborative project involving the NRL (USA), RAL (UK), 
and the Universities of Florence (Italy) and Cambridge (UK).






\begin{figure}
\plotone{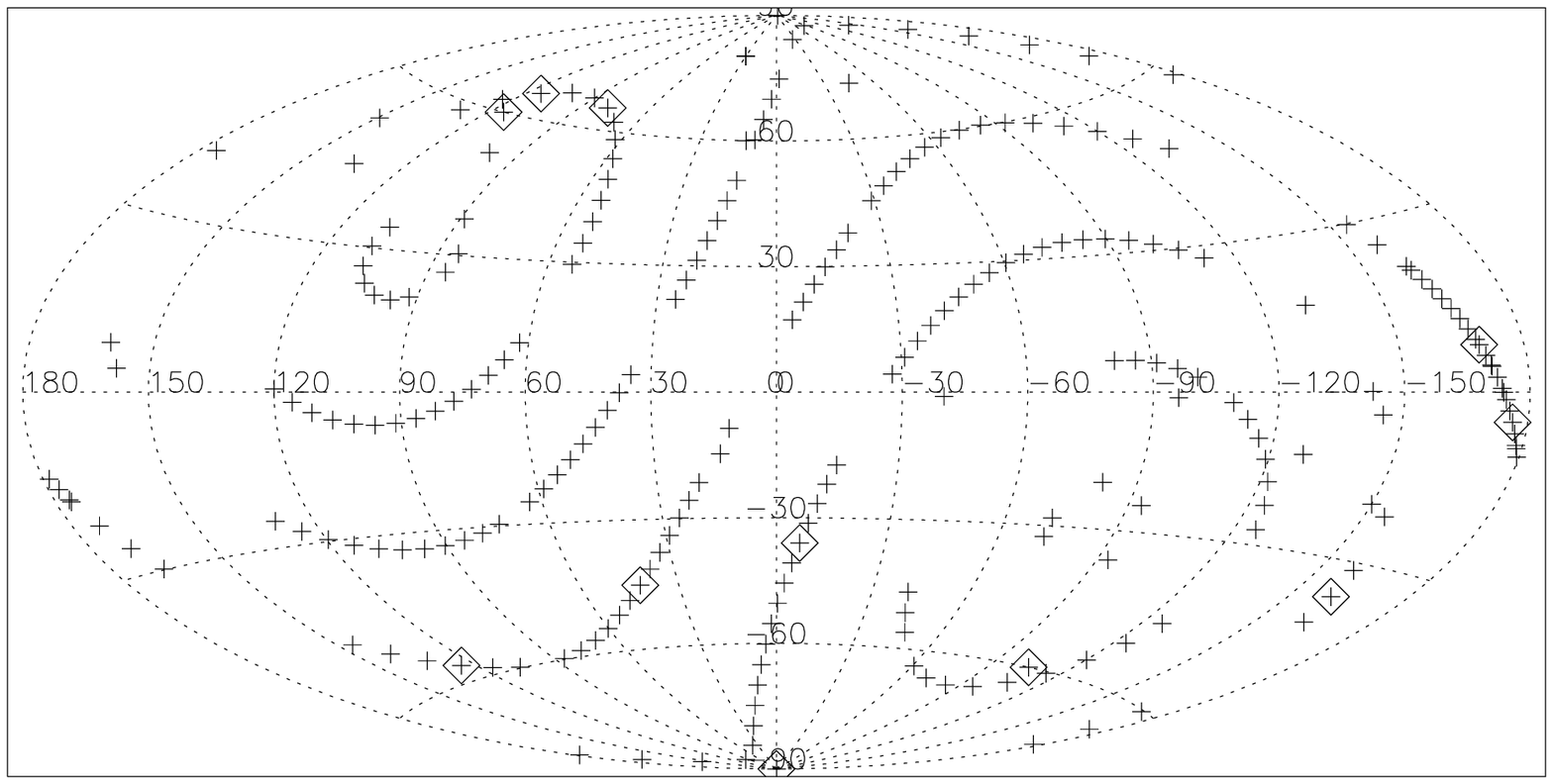}
\caption{Galactic coordinates of CHIPS boresight for the fields 
analyzed in this work.  Large diamonds
indicate targets where the brightest EUV flux was observed by CHIPS.  The
long axis of the CHIPS field of view spans 25 degrees and is generally 
parallel to lines of ecliptic longitude and perpendicular to the 
strings of boresight symbols; thus, the sky coverage has minimal gaps 
when targets are adjacent.  \label{fig1}}
\end{figure}


\clearpage 

\begin{figure}
\plotone{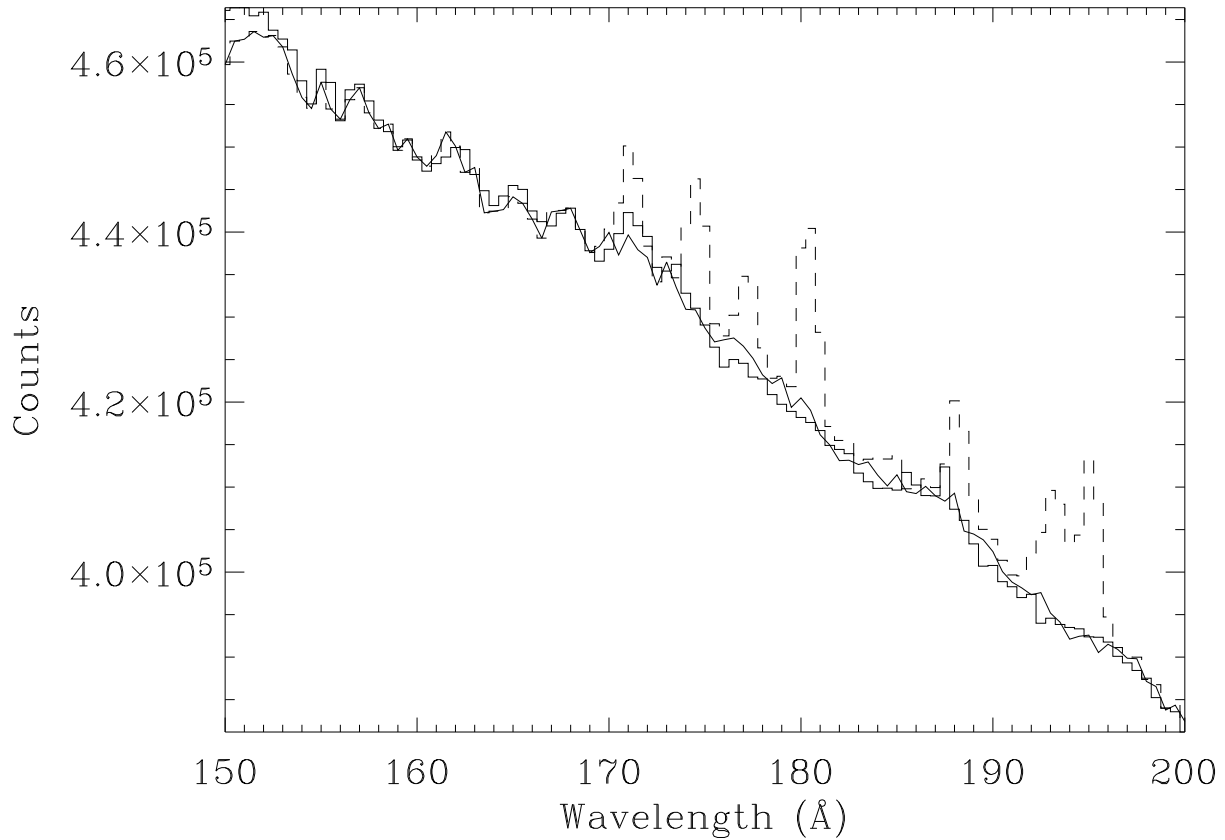}
\caption{Coadded spectrum from 150 - 200 \AA\/ of all fields,
summing both filter panels (solid histogram).  Vertical axis shows total
counts in bins of 0.5 \AA\/.  We also show a scaled,
charged-particle background spectrum (solid line) and
background plus predictions of "canonical" local hot bubble model with
temperature of 10$^{6.1}$ K and emission measure of 0.0038 cm$^{-6}$ pc (dashed histogram),
adapted to the same spectral bin size.
The most significant excess observed above background is near 171 \AA\/. \label{fig2}}
\end{figure}

\clearpage 

\begin{figure}
\plotone{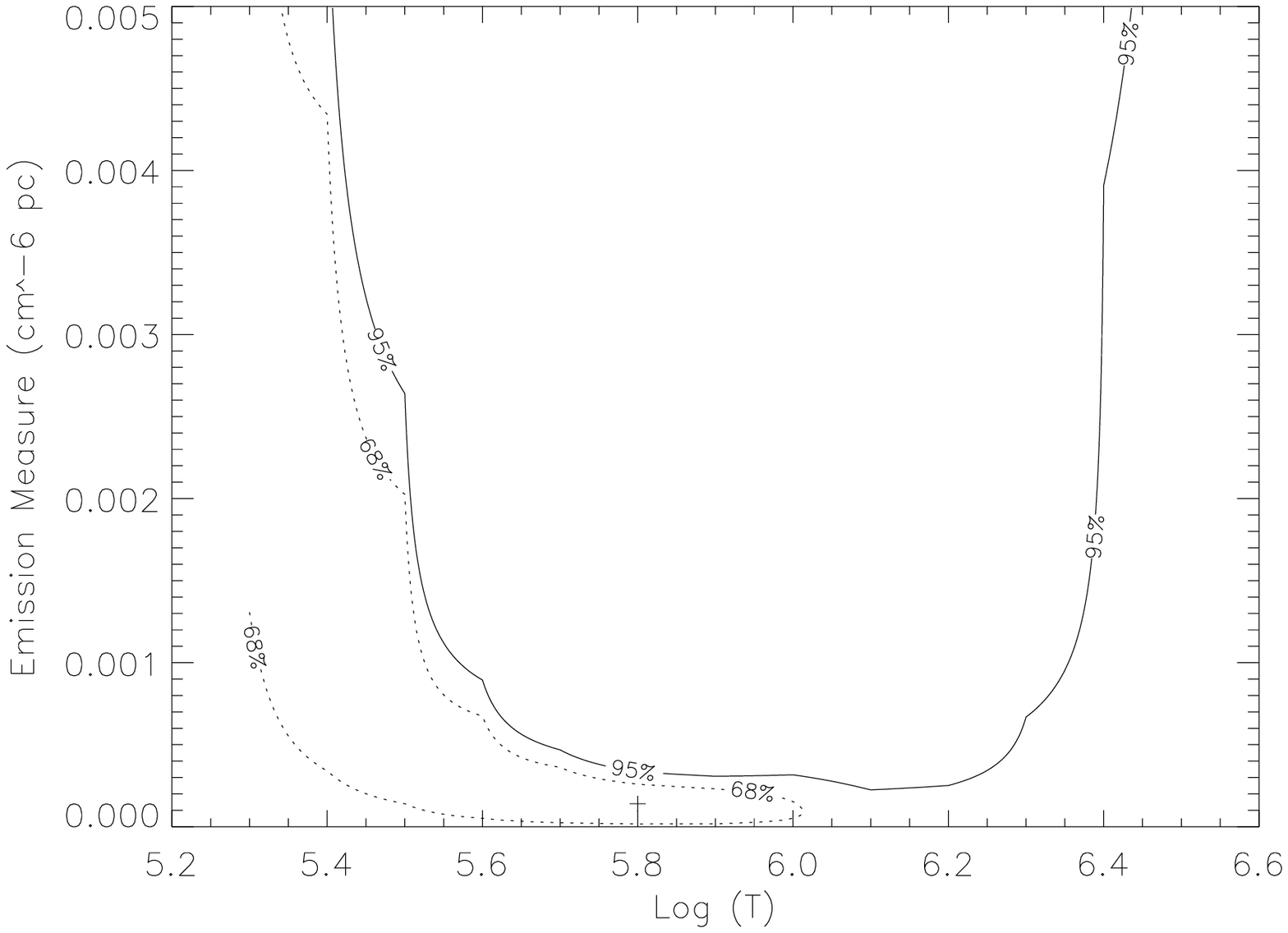}
\caption{Constraints on temperature and emission measure based
on CHIPS observations, and APEC/CHIANTI equilibrium
ionization plasma code with solar abundances.  See text for discussion. \label{fig3}}
\end{figure}

\clearpage 

\begin{figure}
\plotone{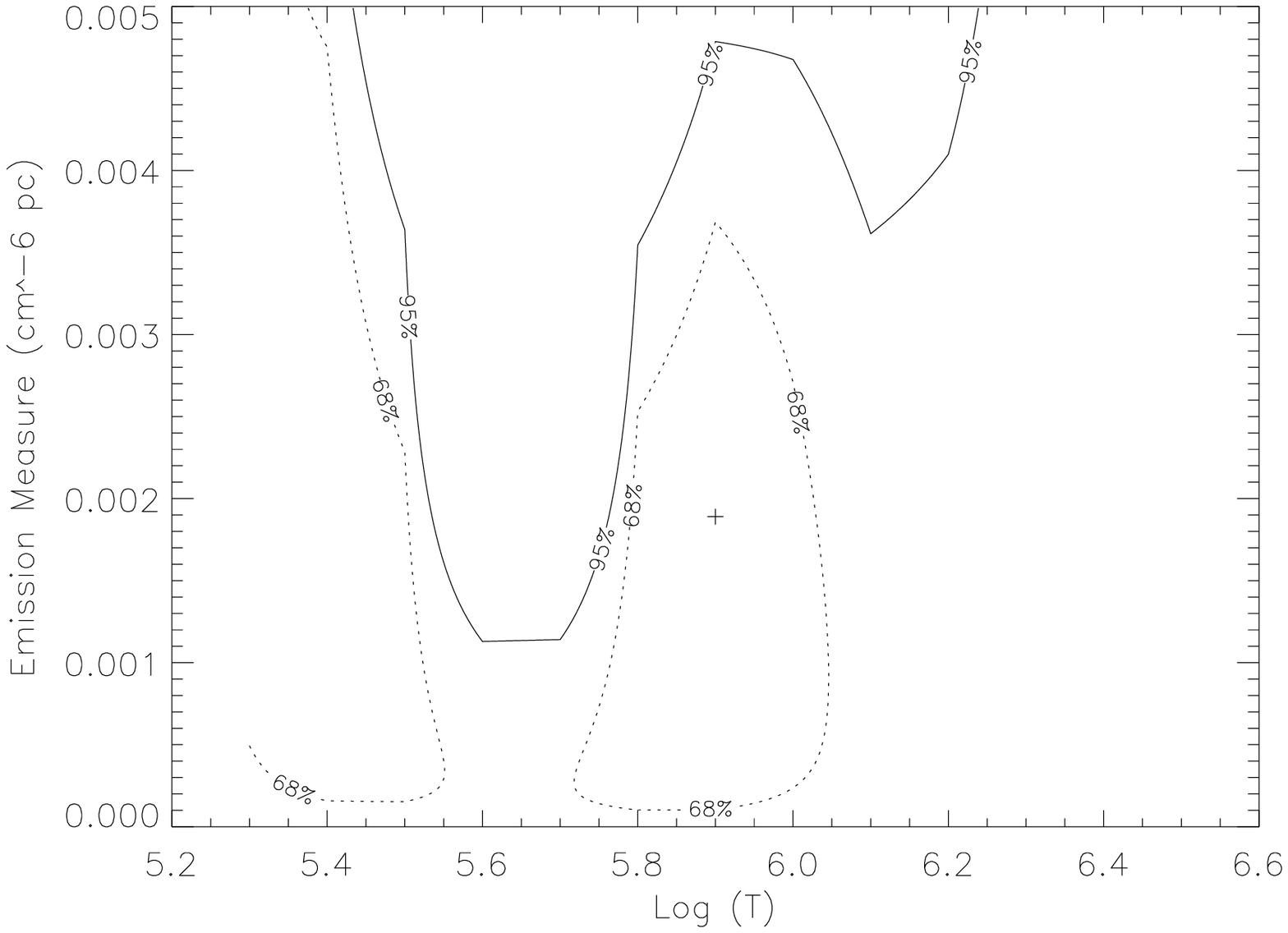}
\caption{Constraints on temperature and emission measure based
on CHIPS observations, and APEC/CHIANTI equilibrium
ionization plasma code with depleted "warm cloud" abundances. 
See text for discussion. \label{fig4}}
\end{figure}

\clearpage
\begin{deluxetable}{llllllll}
\rotate
\tabletypesize{\scriptsize}
\tablecaption{\label{tab:line_limits}}
\tablewidth{0pt}
\tablehead{\colhead{Obs. Wavelength} &
\colhead{Filter(s)} &
\colhead{Flux}   & 
\colhead{Stat. Uncert.} &
\colhead{Bright Flux} & 
\colhead{Stat. Uncert.} &
\colhead{Limiting Uncert.} & 
\colhead{Potential Feature(s)}}
\startdata
116.7&Poly + Zr&0.2&0.9&-0.36&1.8&2.3&Ne VII 116.7 \\
122.6&Poly + Zr&-0.4&0.9&-2.5&1.7&2.2&Ne VI 122.5, 122.7 \\
168.2&Zr&3.4&2.3&-1.2&4.6&5.7&Fe VIII 167.5, 168.2, 168.6, 169.0\\
171.1&Lg. Al + Zr&6.2&1.5&20.&2.9&3.6&Fe IX 171.1\\
174.5&Lg. Al + Zr&2.5&1.6&-2.3&3.1&3.8&Fe X 174.5 \\
177.2&Lg. Al + Zr&-5.4&1.7&-5.6&3.3&4.0&Fe X 177.2 \\
180.4&Lg. Al + Zr&-3.5&1.8&0.1&3.5&4.4&Fe XI 180.4 \\
185.8&Lg. Al&2.8&2.1&0.9&4.1&5.0&Fe VIII 185.2, 186.6 \\
188.3&Lg. Al&-4.0&2.2&0.9&4.3&5.3&Fe XI 188.3 \\
194.5&Lg. Al&-1.9&2.3&0.2&4.4&5.6&Fe XII 193.5, 195.1 \\
202.0&Lg. Al&-0.4&2.6&1.2&4.9&6.1&Fe XIII 202.0 \\
224.7&Lg. Al + Sm. Al&1.1&2.7&5.9&5.1&6.7&S IX 224.7 \\ 
\enddata
{\tt\flushleft{Table 1 notes:
All flux values in LU.  Uncertainties are 1 sigma.
"Flux" refers to flux determined from integrated spectrum
of the full 13.2 Msec of observational data.
"Bright Flux" refers to flux determined from spectrum
formed of the 1.3 Msec when the 171.1 Angstrom feature
is brightest.
"Stat. Uncert." is the uncertainty arising from shot noise
in the spectrum and background subtraction process for
the preceding column; systematic effects are not included.
"Limiting Uncert." is the 1 sigma shot noise for an 800,000 s observation,
and is adopted in the quantitative analysis of the spectra.
"Potential Feature(s)" contains the brightest line(s) expected
from plasma between about 10$^{5.6}$ K  and 10$^{6.2}$ K; at
cooler temperatures, additional features discussed in the text
begin to contribute.
The data generally correspond to two filter panels coadded.  Where
the S/N was higher for a single panel alone, only that panel was used.
At 185.8 \AA\/, the large aluminum filter only was used because of 
a transient "hot spot" on the detector in a region under the zirconium panel.}}

\end{deluxetable}


\clearpage

%
%

%



\begin{thebibliography}{}

\bibitem[Bellm(2003)]{Bellm03} Bellm, E. C. BAAS 2003,111.04.


\bibitem[Bloch(1988)]{B88} Bloch, J.J. 1988, Ph.D. Thesis, Univ. Wisc., Madison.

\bibitem[Bloch et al.(2002)]{Bloch02} Bloch, J.J., Roussel-Dupre, D., Theiler, J., \& Johnson, E. 2002,
Continuing the Challenge of EUV Astronomy: Current Analysis and Prospects for the Future,
ASP Conf. Series, 264, 243, eds. S. Howell, J. Dupuis, D. Golombek, F. Walter \& J. Cullison

\bibitem[Breitschwerdt \& Schmutzler(1994)]{BS94} Breitschwerdt, D. \& Schmutzler, T. 1994, Nature, 371, 774 

\bibitem[Cravens, Robertson, \& Snowden (2001)]{CRS01} Cravens, T.E., Robertson, I.P., \& Snowden, S.L. 2001, \jgr 106, 24, 883. 


\bibitem[Crawford et al.(2002)]{Crayfish02} Crawford, I.A., Lallement,
  R., Price, R. J., Sfeir, D. M., Wakker, P. P., Welsh, B. Y. 2002, \mnras,
  337, 720

\bibitem[Freyberg(1998)]{Frey98} Freyberg, M.J. 1998, in {\it The Local Bubble and Beyond},
Proc. IAU Colloq. No. 166, 113

\bibitem[Hurwitz et al.(2003)]{hurwitz03} Hurwitz, M., et al. 2003 Proc. SPIE 5164, 24.

\bibitem[Janicik et al.(2003)]{janicik03} Janicik, J. et al. 2003 Proc. SPIE 5164, 31.

\bibitem[Jelinsky et al.(1995)]{JVE95} Jelinsky, P.J., Vallerga, J.V., \& Edelstein, J. 1995, \apj, 442, 653

\bibitem[Jenkins et al.(2000)]{Jenkins00} Jenkins, E.B. et al. 2000, \apj, 538L, 81

\bibitem[Juda (1988)]{J88} Juda, M. 1988, Ph.D. Thesis, Univ. Wisc., Madison, 130 pp.

\bibitem[Kimble et al.(1993)]{Kimble93} Kimble, R.A. et al. 1993, \apj, 404, 663

\bibitem[Kuntz \& Snowden(2000)]{KS00} Kuntz, K.D., Snowden, S.L. 2000, \apj,
  543, 195


\bibitem[Lallement et al.(2004)]{lal04} Lallement, R. et al. 2004, in preparation.
\bibitem[Lampton et al.(1976)]{LMB76} Lampton, M., Margon, B., \& Bowyer, S. 1976, \apj, 208, 177

\bibitem[Marckwordt et al.(2003)]{marckwordt03} Marckwordt, M. et al. 2003 Proc. SPIE 5164, 43.

\bibitem[McCammon et al.(2002)]{MCC02} McCammon, D. et al. 2002, \apj, 576, 188.

\bibitem[McCammon \& Sanders(1990)]{MS90} McCammon, D. \& Sanders, W. T. 1990, \araa, 28, 657.

\bibitem[Raymond \& Smith(1977)]{RS77} Raymond, J.C. \& Smith, B.W. 1977, \apjs, 35, 419


\bibitem[Sanders et al.(2001)]{Setal01} Sanders, W.T. et al. 2001, \apj, 554, 694

\bibitem[Savage \& Sembach(1996)]{SS96} Savage, B.D. \& Sembach, K.R. 1996, \araa, 34, 279

\bibitem[Schwadron et al.(1999)]{SFZ99} Schwadron, N.A., Fisk, L.A., \& Zurbuchen, T. H. 1999, \apj, 521, 868.


\bibitem[Sfeir et al.(1999)]{Sfeir99} Sfeir, D.M., Lallement, R., Crifo, F.,
  Welsh, B. Y. 1999, \aap, 346, 785


\bibitem[Sirk et al.(2003)]{sirk03} Sirk, M. M., et al. 2003 Proc. SPIE 5164, 54.

\bibitem[Sholl et al.(2003)]{sholl03} Sholl, M., et al. 2003 Proc. SPIE 5164, 63.

\bibitem[Snowden et al.(1998)]{Sno98} Snowden, S. L., Egger, R., 
  Finkbeiner, D. P., Freyberg, J. J., \& Plucinsky, P. P. 1998, \apj,
  493, 715
\bibitem[Snowden et al.(1997)]{Sno97} Snowden, S. L., Egger, R., 
  Freyberg, M. J., McCammon, D., Plucinsky, P.P, Sanders, W. T., Schmitt,
J. H. M. M., Trumper, J. \& Voges, W. 1997, \apj, 485, 125
\bibitem[Snowden et al.(1993)]{Sno93} Snowden, S. L. et al. 1993,
\apj, 404, 403
Proc. IAU Colloq. No. 166, 103

\bibitem[Vallerga \& Welsh(1995)]{VW95} Vallerga, J.V. \& Welsh, B. Y. 1995, \apj, 444, 702

\bibitem[von Steiger(1994)]{vS94} von Steiger, R., Habilitation Thesis, Univ. of Bern, Switzerland, 1994. 

\bibitem[Welsh et al.(2002)]{Welsh02} Welsh, B. Y., Sallmen, S.,
  Sfeir, D., Shelton, R.I. \& Lallement, R. 2002, \aap, 394, 691
\bibitem[Welsh et al.(1999)]{Welsh99} Welsh, B. Y., Sfeir, D. M., Sirk,
  M. M., Lallement, R. 1999, \aap, 352, 308

\bibitem[Young et al. (2003)]{Yetal03} Young, P.R., et al. 2003, \apjs, 144, 135

\end{thebibliography}
\end{document}